\documentclass[twocolumn,preprintnumbers,showpacs,amsmath,amssymb]{revtex4}
\usepackage{docs}

\usepackage{graphicx}
\usepackage{dcolumn}

\usepackage{bm}

\begin{document}
\title{Chaplygin traversable wormholes}%


\author{Francisco S. N. Lobo}%
\email{flobo@cosmo.fis.fc.ul.pt} \affiliation{Centro de Astronomia
e Astrof\'{\i}sica da Universidade de Lisboa,\\
Campo Grande, Ed. C8 1749-016 Lisboa, Portugal}

\begin{abstract}

The generalized Chaplygin gas (GCG) is a candidate for the
unification of dark energy and dark matter, and is parametrized by
an exotic equation of state given by
$p_{ch}=-A/\rho_{ch}^{\alpha}$, where $A$ is a positive constant
and $0<\alpha \leq 1$. In this paper, exact solutions of
spherically symmetric traversable wormholes supported by the GCG
are found, possibly arising from a density fluctuation in the GCG
cosmological background. To be a solution of a wormhole, the GCG
equation of state imposes the following generic restriction
$A<(8\pi r_0^2)^{-(1+\alpha)}$, where $r_0$ is the wormhole throat
radius, consequently violating the null energy condition. The
spatial distribution of the exotic GCG is restricted to the throat
neighborhood, and the physical properties and characteristics of
these Chaplygin wormholes are further analyzed. Four specific
solutions are explored in some detail, namely, that of a constant
redshift function, a specific choice for the form function, a
constant energy density, and finally, isotropic pressure Chaplygin
wormhole geometries.

\end{abstract}


\maketitle

\section{Introduction}

The nature of the energy content of the Universe is a fundamental
issue in cosmology, and a growing amount of observational evidence
currently favors an accelerating flat Friedmann-Robertson-Walker
model, constituted of $\sim 1/3$ of baryonic and dark matter and
$\sim 2/3$ of a negative pressure dark energy component. The dark
matter content was originally inferred from spiral galactic
rotation curves, which showed a behavior that is significantly
different from the predictions of Newtonian mechanics, and was
later used to address the issue of large scale structure
formation. On the other hand, it has been suggested that dark
energy is a possible candidate for the present accelerated cosmic
expansion \cite{accel-universe}. The dark energy models are
parametrized by an equation of state $\omega=p/\rho<-1/3$, where
$p$ is the spatially homogeneous negative pressure and $\rho$ is
the dark energy density. The range for which $\omega<-1$ has been
denoted phantom energy, and possesses peculiar properties, such
as, an infinitely increasing energy density \cite{Cald}, resulting
in a ``Big Rip'', negative temperatures \cite{Odintsov}, and the
violation of the null energy condition, thus providing a natural
scenario for the existence of wormholes
\cite{Lobo-phantom,phantomWH,stable-phanWH}. In fact, recent fits
to supernovae, cosmic microwave background radiation and weak
gravitational lensing data favor an equation of state with the
dark energy parameter crossing the phantom divide $\omega=-1$
\cite{Vikman,phantom-divide}. Note that $\omega=-1$ corresponds to
the presence of a cosmological constant. It has also been shown,
in a cosmological setting, that the transition into the phantom
regime, for a single scalar field \cite{Vikman} is probably
physically implausible, so that a mixture of various interacting
non-ideal fluids is necessary.

An alternative model is that of the Chaplygin gas, also denoted as
quartessence, based on a negative pressure fluid, which is
inversely proportional to the energy density
\cite{Kamen,GCGbrane2}. The equation of state representing the
generalized Chaplygin gas (GCG) is given by
\begin{equation}
p_{ch}=-\frac{A}{\rho_{ch}^{\alpha}} \,,
\end{equation}
where $A$ and $\alpha$ are positive constants, and the latter lies
in the range $0< \alpha \leq 1$. The particular case of $\alpha=1$
corresponds to the Chaplygin gas. Within the framework of a flat
Friedmann-Robertson-Walker cosmology the GCG equation of state,
after being inserted in the energy conservation equation,
$\dot{\rho}=-3\dot{a}(\rho+p)/a$, yields the following evolution
of the energy density
\begin{equation}
\rho_{ch}=\left[A+\frac{B}{a^{3(1+\alpha)}}\right]^{1/(1+\alpha)}\,,
   \label{rho-evol}
\end{equation}
where $a$ is the scale factor, and $B$ is normally considered to
be a positive integration constant, as to ensure the dominant
energy condition.
However, it is also possible to consider $B<0$, consequently
violating the dominant energy condition, and one verifies that the
energy density will be an increasing function of the scale
function \cite{Lopez-Madrid}.
An attractive feature of this model, is that at early times, the
energy density behaves as matter, $\rho_{ch}\sim a^{-3}$, and as a
cosmological constant at a later stage, $\rho_{ch}={\rm const}$.
It has also been suggested that these two stages are intermediated
by a phase described by a mixture of vacuum energy density and a
soft matter equation of state $p=\alpha \rho$ with $\alpha \neq 1$
\cite{GCGbrane2}. This dual behavior is responsible for the
interpretation that the GCG model is a candidate of a unified
model of dark matter and dark energy \cite{Bilic}, and probably
contains some of the key ingredients in the dynamics of the
Universe for early and late times. In Ref. \cite{Zhang-Wu}, a new
model for describing the unification of dark energy and dark
matter was proposed, which further generalizes the GCG model, and
was thus dubbed the new generalized Chaplygin gas (NGCG) model.
The equation of state of the NGCG is given by $p = -
\tilde{A}(a)/\rho^{\alpha}$, where $a$ is the scale factor and
$\tilde{A}(a)=-wAa^{-3(1+w)(1+\alpha)}$, and the interaction
between dark energy and dark matter is characterized by the
constant $\alpha$.

The GCG model has been confronted successfully with a wide variety
of observational tests, namely, supernovae data \cite{supernovae},
cosmic microwave background radiation \cite{CMB}, gravitational
lensing \cite{gravlens}, gamma-ray bursts \cite{grg} and other
observational data \cite{observational}, which have placed
constraints on the free parameters. A variable Chaplygin gas
model, with $p=-A(a)/\rho$, where $A(a)$ is a positive function of
the scale factor has also been analyzed, and it was found to be
consistent with several observational data for a broad range of
parameters \cite{GuoZhang}. In the context of the NGCG model, it
was shown that the analysis of the observational data also
provides tight constraints on the parameters of the model
\cite{Zhang-Wu}.
The GCG scenario has also been analyzed in some detail in a
modified gravity approach \cite{BarSen}, and another interesting
aspect of the Chaplygin gas is its connection with branes in the
context of string theory. The Chaplygin gas equation of state,
with $\alpha=1$, is associated with the parametrization invariant
Nambu-Goto $d$-brane action in a $d+2$ spacetime
\cite{GCGbrane1,GCGbrane2}, which leads to the action of a
Newtonian fluid, in the light-cone parametrization. Thus, in this
context, the Chaplygin gas corresponds to a gas of $d$-branes in a
$d+2$ spacetime.

It has been argued that a flaw exists in the GCG model, as it
produces oscillations or an exponential blow-up of the matter
power spectrum, which is inconsistent with observations. However,
it has been counter-argued \cite{BBS} that due to the fact that
the GCG is a unique mixture of an interacting dark matter
component and a cosmological dark energy, a flow of energy exists
from the former to the latter. This energy flow is vanishingly
small at early times, but became significant only recently,
leading to a dominance of the dark energy component. It was also
shown that the epoch of the dark energy dominance occurs when the
dark matter perturbations start deviating from its linear
behavior, and that the Newtonian equations for small scale
perturbations for dark matter do not involve any $k$-dependent
term. Therefore, it was concluded that neither oscillations nor a
blow-up in the power spectrum should develop \cite{BBS}.

As emphasized above, recent fits to observational data favor an
evolving equation of state with $\omega$ crossing the phantom
divide $-1$. If confirmed in the future, this behavior holds
important implications to the model construction of dark energy,
and thus excludes the cosmological constant and models with a
constant parameter. In this context, note that the Chaplygin gas
in the dark energy limit cannot cross the phantom divide, however
in Ref. \cite{ZhangZhu} it was shown that an interaction term
between the Chaplygin gas, which in this model plays the role of
dark energy, and dark matter can achieve the phantom crossing. In
Ref. \cite{Meng}, by considering an extension of the Chaplygin
gas, it was shown that the phantom divide $w=-1$ can also be
realized. A further generalization of the GCG was also carried out
in Ref. \cite{Sen-Scher} to allow for the case where $w$ lies in
the phantom regime.

The phantom regime is rather significant, as it violates the null
energy condition (NEC), providing a natural scenario for the
existence of traversable wormholes
\cite{Lobo-phantom,phantomWH,stable-phanWH}.
An interesting feature is that due to the fact of the accelerated
expansion of the Universe, macroscopic wormholes could naturally
be grown from the submicroscopic constructions that originally
pervaded the quantum foam.
In Ref. \cite{gonzalez2} the evolution of wormholes and ringholes
embedded in a background accelerating Universe driven by dark
energy, was analyzed. An interesting feature is that the
wormhole's size increases by a factor which is proportional to the
scale factor of the Universe, and still increases significantly if
the cosmic expansion is driven by phantom energy. The accretion of
dark and phantom energy onto Morris-Thorne
wormholes~\cite{diaz-phantom3,diaz-phantom4}, was further
explored, and it was shown that this accretion gradually increases
the wormhole throat which eventually overtakes the accelerated
expansion of the universe, consequently engulfing the entire
Universe, and becomes infinite at a time in the future before the
big rip. This process was dubbed the ``Big Trip''
\cite{diaz-phantom3,diaz-phantom4}. It was shown that using
$k-$essence dark energy also leads to the big rip \cite{PGDiaz-k},
although, in an interesting article \cite{diaz-phantom},
considering a generalized Chaplygin gas the big rip may be avoided
altogether.
In this paper, we shall be primarily interested in considering the
possibility that traversable wormholes may be supported by the GCG
equation of state. In this context, and related to the features of
the Big Trip, the accretion of a generalized Chaplygin gas onto
wormholes was explored in Ref. \cite{Madrid}. Several cases were
extensively analyzed. Imposing the dominant energy condition, it
was found that the evolution of the wormhole mass decreases with
cosmic time. Considering the violation of the dominant energy
condition, i.e., with $B<0$, as the wormhole accretes Chaplygin
phantom energy, the wormhole mass increases from an initial value,
and reaches a plateau as time tends to infinity. In fact, a wide
region of the Chaplygin parameters were found where the Big Trip
is avoided.

Despite the fact that the GCG in the dark energy regime describes
a spatially homogeneously distributed fluid, it has been pointed
out that the GCG equation of state is that of a polytropic gas
with a negative polytropic index \cite{BertPar}, and thus
inhomogeneous structures, such as a GCG dark energy star may arise
from a density fluctuation in the GCG cosmological background.
Other astrophysical implications of the model have also been
analyzed. In Ref. \cite{BTV}, a gravitational vacuum star solution
(gravastar) was constructed by replacing the interior de Sitter
solution, with the Chaplygin gas equation of state in the phantom
regime, which was dubbed a Born-Infeld phantom gravastar. A
generalization of the gravastar picture (see Ref.
\cite{stable-darkstar} and references therein), with the inclusion
of an interior solution governed by the equation of state
$\omega=p/\rho<-1/3$ was also analyzed in Ref.
\cite{stable-darkstar}.

In this work, we shall construct static and spherically symmetric
traversable wormhole geometries, satisfying the GCG equation of
state, which we denote ``Chaplygin wormholes'', by considering a
matching of these geometries to an exterior vacuum spacetime, and
further analyze the physical properties and characteristics of
these solutions. We find that the spatial distribution of the
exotic GCG is restricted to the throat neighborhood. We shall also
consider specific solutions, and explore the traversability
conditions \cite{Morris,Visser} and apply the ``volume integral
quantifier'' \cite{VKD1}, which amounts to measuring the amount of
averaged null energy condition violating matter, to particular
cases.

This paper is outlined in the following manner. In Sec. II, we
present a general solution of a traversable wormhole supported by
a generalized Chaplygin gas, with a cut-off of the stress-energy
tensor at a junction interface. In Sec. III, specific wormhole
geometries are analyzed and several of their physical properties
and characteristics are explored in some detail, namely, that of a
constant redshift function, a specific choice for the form
function, a constant energy density, and finally, isotropic
pressure Chaplygin wormhole geometries. Finally, in Sec. IV, we
conclude.

\section{Chaplygin wormholes}

\subsection{Metric and field equations}

The spacetime metric representing a spherically symmetric and
static wormhole is given by
\begin{equation}
ds^2=-e ^{2\Phi(r)}\,dt^2+\frac{dr^2}{1- b(r)/r}+r^2 \,(d\theta
^2+\sin ^2{\theta} \, d\phi ^2) \label{metricwormhole}\,,
\end{equation}
where $\Phi(r)$ and $b(r)$ are arbitrary functions of the radial
coordinate, $r$, denoted as the redshift function, and the form
function, respectively \cite{Morris}. The radial coordinate has a
range that increases from a minimum value at $r_0$, corresponding
to the wormhole throat, to infinity. One may also consider a
cut-off of the stress-energy tensor at a junction radius $a$.

A fundamental property of a wormhole is that a flaring out
condition of the throat, given by $(b-b'r)/b^2>0$, is imposed
\cite{Morris,Visser}. The latter may be deduced from the
mathematics of embedding, and from this we verify that at the
throat $b(r_0)=r=r_0$, the condition $b'(r_0)<1$ is imposed to
have wormhole solutions. Another condition that needs to be
satisfied is $1-b(r)/r>0$, i.e., $b(r)<r$. For the wormhole to be
traversable, one must demand that there are no horizons present,
which are identified as the surfaces with $e^{2\Phi}\rightarrow
0$, so that $\Phi(r)$ must be finite everywhere.

Using the Einstein field equation, $G_{\mu\nu}=8\pi \,T_{\mu\nu}$,
(with $c=G=1$) we obtain the following relationships
\begin{eqnarray}
b'&=&8\pi r^2 \rho  \label{rhoWH}\,,\\
\Phi'&=&\frac{b+8\pi r^3 p_r}{2r^2(1-b/r)}  \label{prWH}\,,\\
p_r'&=&\frac{2}{r}\,(p_t-p_r)-(\rho +p_r)\,\Phi ' \label{ptWH}\,,
\end{eqnarray}
where the prime denotes a derivative with respect to the radial
coordinate, $r$. $\rho(r)$ is the energy density, $p_r(r)$ is the
radial pressure, and $p_t(r)$ is the lateral pressure measured in
the orthogonal direction to the radial direction. Equation
(\ref{ptWH}) may be obtained using the conservation of the
stress-energy tensor, $T^{\mu\nu}_{\;\;\;\;;\nu}=0$, which can be
interpreted as the hydrostatic equation for equilibrium for the
material threading the wormhole.

Another fundamental property of wormholes is the violation of the
null energy condition (NEC), $T_{\mu\nu}k^\mu k^\nu \geq 0$, where
$k^\mu$ is {\it any} null vector \cite{Morris,HochVisser}. From
Eqs. (\ref{rhoWH}) and (\ref{prWH}), considering an orthonormal
reference frame with $k^{\hat{\mu}}=(1,1,0,0)$, so that
$T_{\hat{\mu}\hat{\nu}}k^{\hat{\mu}} k^{\hat{\nu}}=\rho+p_r$, one
verifies
\begin{equation}
\rho(r)+p_r(r)=\frac{1}{8\pi}\,\left[\frac{b'r-b}{r^3}+
2\left(1-\frac{b}{r}\right) \frac{\Phi '}{r} \right]  \,.
\end{equation}
Evaluated at the throat, $r_0$, and considering the flaring out
condition and the finite character of $\Phi(r)$, we have
$\rho+p_r<0$. Matter that violates the NEC is denoted {\it exotic
matter}.

\subsection{GCG equation of state}

The equation of state representing the generalized Chaplygin gas
(GCG) is given by $p_{ch}=-A/\rho_{ch}^{\alpha}$, where $A$ and
$\alpha$ are positive constants, and the latter lies in the range
$0< \alpha \leq 1$. The particular case of $\alpha=1$ corresponds
to the Chaplygin gas. An attractive feature of this model, as
mentioned in the Introduction, is that at early times, the energy
density behaves as matter, $\rho_{ch}\sim a^{-3}$, and as a
cosmological constant at a later stage, $\rho_{ch}={\rm const}$.
In a cosmological context, at a late stage dominated by an
accelerated expansion of the Universe, the cosmological constant
may be given by $8\pi A^{1/(1+\alpha)}$. This dual behavior is
responsible for the interpretation that the GCG model is a
candidate of a unified model of dark matter and dark energy. It
has been shown that GCG can be algebraically decomposed into a
dark matter and a dark energy component \cite{BBS}, in which there
exists a transference of energy from the former to the latter.

It was noted in Ref. \cite{BertPar} that the GCG equation of state
is that of a polytropic gas with a negative polytropic index, and
thus suggested that one could analyze astrophysical implications
of the model. In this context, we shall explore the construction
of traversable wormholes, possibly from a density fluctuation in
the GCG cosmological background. As in Refs.
\cite{phantomWH,Lobo-phantom}, we will consider that the pressure
in the GCG equation of state is a radial pressure, and the
tangential pressure can be determined from the Einstein equations,
namely, Eq. (\ref{ptWH}). Thus, taking into account the GCG
equation of state in the form $p_r=-A/\rho^{\alpha}$, and using
Eqs. (\ref{rhoWH})-(\ref{prWH}), we have the following condition
\begin{equation}
\Phi'(r)=\left[-A(8\pi)^{1+\alpha}\;\frac{r^{2\alpha +1}
}{2(b')^{\alpha}}+\frac{b}{2r^2}\right]\Big/\left(1-\frac{b}{r}
\right) \,.
            \label{EOScondition}
\end{equation}
Solutions of the metric (\ref{metricwormhole}), satisfying Eq.
(\ref{EOScondition}) shall be denoted ``Chaplygin wormholes''.

We now have a system of four equations, namely, Eqs.
(\ref{rhoWH})-(\ref{ptWH}) and Eq. (\ref{EOScondition}), with five
unknown functions of $r$, i.e., the stress-energy components,
$\rho(r)$, $p_r(r)$ and $p_t(r)$, and the metric fields, $b(r)$
and $\Phi(r)$. To construct specific solutions, we may adopt
several approaches, and in this work we shall mainly use the
strategy of considering restricted choices for $b(r)$ and
$\Phi(r)$, in order to obtain solutions with the properties and
characteristics of wormholes. One may also impose a specific form
for the stress-energy components and through the field equations
and Eq. (\ref{EOScondition}) determine $b(r)$ and $\Phi(r)$.
Throughout this paper, we shall consider the cases that the energy
density is positive $\rho>0$, which implies that only form
functions of the type $b'(r)>0$ are considered.

As shown above, to be a wormhole solution, the condition
$b'(r_0)<1$ is imposed. Now, using the GCG equation of state,
evaluated at the throat, and taking into account Eq. (\ref{prWH}),
we verify that the energy density at $r_0$ is given by
$\rho(r_0)=\left(8\pi r_0^2 A\right)^{1/\alpha}$. Finally, using
Eq. (\ref{rhoWH}), and the condition $b'(r_0)<1$, we verify that
for Chaplygin wormholes, the following condition is imposed
\begin{equation}
A< \left(8\pi r_0^2\right)^{-(1+\alpha)}\,.
    \label{ChapWH-restrict}
\end{equation}
It is a simple matter to show that this condition necessarily
violates the NEC at the wormhole throat. However, for the GCG
cosmological models it is generally assumed that the NEC is
satisfied, i.e, $p+\rho \geq 0$, which implies $\rho \geq
A^{1/(1+\alpha)}$. The NEC violation is a fundamental ingredient
in wormhole physics, and it is in this context that we shall
explore the construction of traversable wormholes, i.e., for $\rho
< A^{1/(1+\alpha)}$. Note that as emphasized in Refs.
\cite{Lopez-Madrid,Madrid}, considering a negative integration
constant, $B<0$, in the evolution of the energy density, Eq.
(\ref{rho-evol}), one also deduces that $\rho_{ch} <
A^{1/(1+\alpha)}$. This condition violates the dominant energy
condition, and is consistent with the analysis outlined in this
paper, proving the compatibility of both works.

Note that the velocity of sound has been interpreted as
$v_s^2=\partial p/\partial \rho=A\alpha/\rho^{1+\alpha}$. Thus,
from the condition that the latter should not exceed the speed of
light, i.e., $v_s\leq 1$, and from the violation of the NEC,
$\rho+p<0$ we have the additional constraints
$1<A/\rho^{1+\alpha}\leq 1/\alpha$. The latter evaluated at the
throat, takes the form $\alpha^{\alpha}\leq A (8\pi
r_0^2)^{(1+\alpha)}<1$, for $\alpha<1$. However, it is worth
pointing out that in the presence of exotic matter, one cannot
naively interpret $\partial p/\partial \rho$ as the speed of
sound, as a detailed microphysical model describing the physics of
exotic matter is still lacking. Therefore, one cannot {\it a
priori} impose $0<\partial p/\partial \rho\leq 1$, and it is worth
noting that there are several known examples of exotic $\partial
p/\partial \rho<0$ behavior, namely the Casimir effect and the
false vacuum. (See Ref. \cite{Poisson} for a detailed analysis).

\subsection{Stress-energy tensor cut-off}

We can construct asymptotically flat spacetimes, in which
$b(r)/r\rightarrow 0$ and $\Phi\rightarrow 0$ as $r\rightarrow
\infty$. However, one may also construct solutions with a cut-off
of the stress-energy, by matching the interior solution of metric
(\ref{metricwormhole}) to an exterior vacuum spacetime, at a
junction interface. If the junction contains surface stresses, we
have a thin shell, and if no surface stresses are present, the
junction interface is denoted a boundary surface. The solutions
analyzed in this work are not asymptotically flat, where the
spatial distribution of the exotic GCG is restricted to the throat
neighborhood, so that the dimensions of these Chaplygin wormholes
are not arbitrarily large.

For simplicity, consider that the exterior vacuum solution is the
Schwarzschild spacetime, given by the following metric
\begin{equation}
ds^2=-\left(1-\frac{2M}{r}\right)\,dt^2+\frac{dr^2}{1-
\frac{2M}{r}}+r^2 \,(d\theta ^2+\sin ^2{\theta} \, d\phi ^2)
\label{Sch-metric}.
\end{equation}
Note that the matching occurs at a radius greater than the event
horizon $r_b=2M$, i.e., $a>2M$. The Darmois-Israel formalism
\cite{Darmois-Israel} then provides the following expressions for
the surface stresses of a dynamic thin shell
\cite{stable-phanWH,LoboCrawford}
\begin{eqnarray}
\sigma&=&-\frac{1}{4\pi a} \left(\sqrt{1-\frac{2M}{a}+\dot{a}^2}-
\sqrt{1-\frac{b(a)}{a}+\dot{a}^2} \, \right)
    \label{surfenergy}   ,\\
{\cal P}&=&\frac{1}{8\pi a} \Bigg[\frac{1-\frac{M}{a}
+\dot{a}^2+a\ddot{a}}{\sqrt{1-\frac{2M}{a}+\dot{a}^2}}
   \nonumber    \\
&&-\frac{(1+a\Phi') \left(1-\frac{b}{a}+\dot{a}^2
\right)+a\ddot{a}-\frac{\dot{a}^2(b-b'a)}{2(a-b)}}{\sqrt{1-\frac{b(a)}{a}+\dot{a}^2}}
\, \Bigg]         \,,
    \label{surfpressure}
\end{eqnarray}
where the overdot denotes a derivative with respect to the proper
time, $\tau$. $\sigma$ and ${\cal P}$ are the surface energy
density and the tangential surface pressure, respectively. The
static case is given by taking into account $\dot{a}=\ddot{a}=0$
\cite{wormhole-shell}.

\section{Specific solutions}

\subsection{Constant redshift function}

Consider, for instance, a constant redshift function,
$\Phi'(r)=0$, so that from Eq. (\ref{EOScondition}), one
determines the following form function
\begin{eqnarray}
b(r)&=&r_0\Bigg[(3A)^{1/\alpha}\left(\frac{8\pi}{3r_0}\right)
^{(1+\alpha)/\alpha}  \times
    \nonumber    \\
&&\times\left(r^{3(\alpha+1)/\alpha}
-r_0^{3(\alpha+1)/\alpha}\right) +1\Bigg]^{\alpha/(1+\alpha)} \,.
     \label{form3}
\end{eqnarray}
For the particular case of the Chaplygin gas, $\alpha=1$, the form
function reduces to
\begin{eqnarray}
b(r)=r_0\left[\frac{64}{3}\frac{A\pi^2}{r_0^2}(r^6-r_0^6)+1\right]^{1/2}
\,.
     \label{form3b}
\end{eqnarray}

\begin{figure}[h]
\centering
  \includegraphics[width=2.8in]{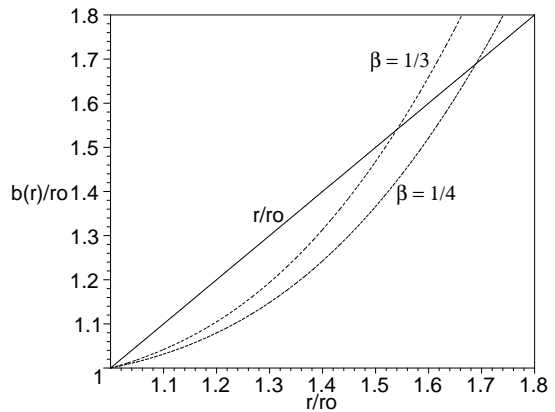}
  \caption{To be a solution of a wormhole, the condition $b(r)<r$ is
  imposed, so that only the interval below the solid line $r/r_0$
  provide wormhole solutions.
  The plot depicts the values of $\beta=1/3$ and $\beta=1/4$, and
  one verifies that the wormhole dimensions increase for decreasing
  values of $\beta$. See the text for details.}
  \label{Fig:form}
\end{figure}

To be a solution of a wormhole, the condition $b'(r_0)<1$ is
imposed. Thus, from the latter condition and Eq. (\ref{form3b}),
we deduce the restriction $A<(8\pi r_0^2 )^{-2}$. For instance,
considering $A=\beta(8\pi r_0^2)^{-2}$, with $0<\beta<1$, the form
function is given by
\begin{eqnarray}
b(r)=r_0\left\{\frac{\beta}{3}\left[\left(\frac{r}{r_0}\right)^6-1\right]+1\right\}^{1/2}
\,.
\end{eqnarray}
Note that this does not correspond to an asymptotically flat
spacetime, however, one may match this solution to an exterior
vacuum geometry, so that the dimensions of this specific Chaplygin
wormhole cannot be arbitrarily large. To be a solution of a
wormhole, the condition $b(r)<r$ is also imposed.
Note that $b(r)=r$ has two real and positive roots given by
$r_-=r_0$ and $r_+=r_0\{[(12/\beta -3)^{1/2}-1]/2\}^{1/2}$, so
that to be a solution of a wormhole, $r$ lies in the range
\begin{equation}
r_0<r<r_0\left\{\frac{1}{2}\left[\left(\frac{12}{\beta}
-3\right)^{1/2}-1\right]\right\}^{1/2}  \,.
\end{equation}
This restriction is shown graphically in Fig. \ref{Fig:form} for
the values of $\beta=1/3$ and $\beta=1/4$. Note that the wormhole
dimensions increase for decreasing values of $\beta$.

It is also of interest to consider the traversability conditions
required for a human being to journey through the wormhole. Rather
than reproduce the results here, we refer the reader to Refs.
\cite{Morris,Visser,Lobo-phantom}. We shall consider, for
simplicity, a constant and non-relativistic traversal velocity for
the traveller. Note that for the specific solutions considered in
this subsection, the conditions required that the acceleration
felt by the traveller, and that the radial tidal acceleration
should not exceed Earth's gravity, $g_\oplus$, are readily
satisfied (see Refs. \cite{Morris,Visser,Lobo-phantom} for
details). From the condition that the lateral tidal acceleration
should not exceed Earth's gravitational acceleration, evaluated at
the wormhole throat $r_0$, one obtains the inequality
\begin{equation}
v \leq
r_0\,\sqrt{\frac{2g_\oplus}{(1-\beta)\,\big|\eta^{\hat{2}'}\big|}}\,,
\end{equation}
where $\eta^{\hat{2}'}$ is the separation between two arbitrary
parts of the traveller's body, measured along the lateral
direction in the traveller's reference frame, and, for simplicity,
we shall assume that $|\eta^{\hat{2}'}|\approx 2\,{\rm m}$
\cite{Morris,Lobo-phantom}.

Now, as in Ref. \cite{Lobo-phantom}, considering the equality
case, with $\beta=1/2$, and assuming that the wormhole throat is
given by $r_0\approx 10^2\,{\rm m}$, the traversal velocity takes
the following value $v\approx 4\times 10^2\,{\rm m/s}$.
Considering that the junction radius is given by $a\approx
10^4\,{\rm m}$, then one obtains the traversal times of $\Delta
\tau \approx \Delta t \approx 50 \,{\rm s}$, as measured by the
traveller and for the observers that remain at rest at the space
stations situated at $a$, respectively (see Ref.
\cite{Lobo-phantom} for details). It is interesting to note that
these traversability conditions are identical to the specific case
of an asymptotically flat phantom wormhole spacetime analyzed in
Ref. \cite{Lobo-phantom}.

\subsection{$b(r)=r_0\sqrt{r/r_0}$}

Consider $b(r)=r_0\sqrt{r/r_0}$, so that from Eq.
(\ref{EOScondition}), considering the Chaplygin gas with
$\alpha=1$, one determines the following redshift function
\begin{eqnarray}
\Phi(r)&=&-128A\pi^2 r_0^4 \left[\ln\left(\sqrt{\frac{r}{r_0}}-1
\right)+\sum_{n=1}^9\;\frac{1}{n}\left(\frac{r}{r_0}\right)^{n/2}\right]
    \nonumber    \\
&&+\ln\left(1-\sqrt{\frac{r_0}{r}}\right)  +C
         \label{Phi1}
\end{eqnarray}
The constant of integration, $C$, may be determined from the
boundary conditions, $\Phi(a)$, at the junction interface. Note
that this solution reflects a non-traversable wormhole as it
possesses an event horizon at the throat $r=r_0$, as may be
readily verified from the first term in square brackets in the
right hand side of Eq. (\ref{Phi1}).
However, imposing the condition $A=(128\pi^2 r_0^4)^{-1}$, Eq.
(\ref{Phi1}) reduces to
\begin{eqnarray}
\Phi(r)=-\sum_{n=1}^9\;\frac{1}{n}\left(\frac{r}{r_0}\right)^{n/2}
-\ln\left(\sqrt{\frac{r}{r_0}}\right)  +C  \,.
         \label{Phi2}
\end{eqnarray}
As in the previous example, this solution is not asymptotically
flat, however, one may match the latter to an exterior vacuum
spacetime at a junction radius $a$. Note that the constant $C$ is
given by
\begin{eqnarray}
C=\Phi(a)+\sum_{n=1}^9\;\frac{1}{n}\left(\frac{a}{r_0}\right)^{n/2}
+\ln\left(\sqrt{\frac{a}{r_0}}\right)  \,.
         \label{constant-C}
\end{eqnarray}
This solution now reflects a traversable wormhole, as the redshift
function is finite in the range $r_0\leq r \leq a$.

Using the ``volume integral quantifier'', which provides
information about the ``total amount'' of averaged null energy
condition (ANEC) violating matter in the spacetime (see Refs.
\cite{VKD1} for details), given by
$I_V=\int\left[\rho(r)+p_r(r)\right]dV$, with a cut-off of the
stress-energy at $a$, we have
\begin{eqnarray}
I_V&=&\left[r\left(1-\frac{b}{r}\right)
\ln\left(\frac{e^{2\Phi}}{1-b/r}\right)\right]_{r_0}^a
    \nonumber    \\
&&-\int_{r_0}^a
\left(1-b'\right)\left[\ln\left(\frac{e^{2\Phi}}{1-b/r}\right)\right]\;dr
    \nonumber   \\
&=&\int_{r_0}^a
\left(r-b\right)\left[\ln\left(\frac{e^{2\Phi}}{1-b/r}\right)\right]'\;dr
\,.
    \label{vol-int}
\end{eqnarray}
Now, using the form and redshift functions provided above, and
evaluating the integral, one finally ends up with the following
simplified expression for the ``volume integral quantifier''
\begin{eqnarray}
I_V=r_0\left[\sqrt{\frac{a}{r_0}}-\frac{2}{11}\left(\frac{a}{r_0}\right)^{11/2}-\frac{9}{11}\right]
\,.
\end{eqnarray}
By taking the limit $a\rightarrow r_0$, one readily verifies that
$I_V\rightarrow 0$. This proves that as in the specific case of
phantom wormholes \cite{Lobo-phantom}, one may theoretically
construct a wormhole with arbitrarily small amounts of a Chaplygin
gas. As emphasized in Ref. \cite{Lobo-phantom}, this result is not
unexpected. However, it is interesting to note the relative ease
that one may theoretically construct wormholes supported by
infinitesimal amounts of exotic fluids used in cosmology to
explain the present accelerated cosmic expansion.

\subsection{Constant energy density}

Considering a constant energy density, $\rho=\rho_0$, we verify
from Eq. (\ref{rhoWH}) that the form function is given by
\begin{equation}
b(r)=C \left(r^3-r_0^3 \right) +r_0 \,,
   \label{b-const-rho}
\end{equation}
with the definition $C=8\pi \rho_0/3$. The condition $b'(r_0)<1$,
imposes the following restriction $3Cr_0^2<1$. Consider
$C=\beta(3r_0^2)^{-1}$, with $0<\beta<1$, so that Eq.
(\ref{b-const-rho}) takes the form
\begin{equation}
b(r)=r_0\left\{\frac{\beta}{3}\left[\left(\frac{r}{r_0}\right)^3-1\right]
+1\right\}   \,.
\end{equation}
To be a wormhole solution the condition $b(r)<r$ is also imposed.
Note that $b(r)=r$ has two real positive roots given by $r_-=r_0$
and $r_+=r_0(\sqrt{12/\beta-3}-1)/2$, so that $r$ lies in the
range
\begin{equation}
r_0<r<\frac{r_0}{2}\left(\sqrt{\frac{12}{\beta}-3}-1\right)\,.
    \label{range}
\end{equation}

From Eq. (\ref{EOScondition}), the redshift function is formally
given by
\begin{eqnarray}
&\hspace{-0.5cm}\Phi(r)=-\frac{1}{2}\int
\frac{\beta(r^3-r_0^3)-24\pi A \left(8\pi
r_0^2/\beta\right)^{\alpha}r_0^2
r^3+3r_0}{r(r-r_0)\left[r^2+r_0r-(3-\beta)r_0^2\right]}\;dr
    +  C_1 .
          \label{const-Phi}
\end{eqnarray}
The constant of integration, $C_1$, may be determined from the
boundary conditions, at the junction interface, $a$. For the
particular case of the Chaplygin gas, $\alpha=1$, Eq.
(\ref{const-Phi}) is integrated to provide
\begin{eqnarray}
\Phi(r)&=&-\frac{1}{2\beta(1-\beta)}\Bigg\{\left(64A\pi^2
r_0^4-\beta \right)\,\ln (r-r_0)
      \nonumber  \\
&&\hspace{-1.4cm}+\left[\frac{\beta}{2}-\frac{32A\pi^2
r_0^4(3-2\beta)}{\beta} \right] \ln \left[\beta r^2+\beta r_0
r-(3-\beta) r_0^2 \right]
       \nonumber   \\
&&\hspace{-1.0cm}-\frac{3(\beta^2-64A\pi^2 r_0^4)}{\sqrt
{3\beta(4- \beta)}} \;
{\rm arctanh} \left[\frac {\beta(2r+r_0)}{r_0\sqrt {3\beta(4-
\beta)}}\right]\Bigg\}
     \nonumber    \\
&&-\frac{1}{2}\,\ln (r) +C_1\,.
\end{eqnarray}
One verifies, as in the previous case, that this is a solution of
a non-traversable wormhole, as an event horizon exists at $r_0$.
However, imposing the condition $A=\beta(64\pi^2 r_0^4)^{-1}$ with
$0<\beta<1$, the redshift function reduces to
\begin{eqnarray}
\Phi(r)&=&-\frac{1}{2}\,\ln (r)+\frac{3}{4\beta} \ln \left[\beta
r^2+\beta r_0r -(3-\beta)r_0^2 \right]
       \nonumber   \\
&&\hspace{-1.3cm}+\frac{1}{2\sqrt {3\beta (4-\beta)}} \;{\rm
arctanh} \left[\frac {\beta(2r+r_0)}{r_0\sqrt {3\beta
(4-\beta)}}\right]+C_1
 \,.
     \label{Phi3}
\end{eqnarray}
It is a simple matter to show that $\Phi(r)$ given by Eq.
(\ref{Phi3}) is finite in the range (\ref{range}), so that this
solution now reflects a traversable wormhole. This interior
wormhole solution is now matched to an exterior vacuum spacetime
within the range of the inequality (\ref{range}). Note that the
condition $A=\beta(64\pi^2 r_0^4)^{-1}$ satisfies the restriction
of the inequality (\ref{ChapWH-restrict}).

\subsection{Isotropic pressure}

Consider an isotropic pressure, $p_r=p_t$, so that from Eq.
(\ref{ptWH}), we have the differential equation
\begin{eqnarray}
\frac{A\alpha \rho'}{\rho\left(A-\rho^{\alpha+1}\right)}= \Phi'(r)
\,,
\end{eqnarray}
which has the following solution
\begin{eqnarray}
\rho(r)=\left[A^{-1}+e^{-\frac {\left (\alpha+1\right
)}{\alpha}\Phi (r)}C_1\right ]^{-\frac{1}{1+\alpha}} \,.
    \label{iso-rho}
\end{eqnarray}
From the GCG equation of state, $p_r=-A/\rho^\alpha$, and using
Eq. (\ref{prWH}), we verify that the energy density, evaluated at
the throat, reduces to $\rho(r_0)=\left(8\pi r_0^2
A\right)^{1/\alpha}$, so that the constant $C_1$ is given by
\begin{eqnarray}
C_1=\left[\left(8\pi r_0^2
A\right)^{-(1+\alpha)/\alpha}-A^{-1}\right]e^{\frac {\left
(\alpha+1\right )}{\alpha}\Phi (r_0)} \,.
      \label{C1-constant}
\end{eqnarray}

Taking into account Eq. (\ref{rhoWH}), and considering a specific
choice for the redshift function, for instance,
\begin{equation}
\Phi(r)=\ln\left(\frac{r}{r_0}\right)  \,,
\end{equation}
we have
\begin{equation}
b'(r)=8\pi r^2\left[A^{-1}+C_1\left(\frac{r}{r_0}\right)
^{-(1+\alpha)/\alpha}\right]^{-1/(1+\alpha)} \,.
     \label{formderiv}
\end{equation}
Equation (\ref{formderiv}) may be integrated to provide
\begin{eqnarray}
&b(r)=C_2+(8\pi r^3/3)\;A^{1/(1+\alpha)} \times
        \nonumber \\
&\hspace{-0.5cm}\times{\rm
hypergeom}\left(\left[\frac{1}{1+\alpha},\frac{-3\alpha}{1+\alpha}\right],
\left[\frac{1-2\alpha}{1+\alpha}\right],-\left(\frac{r_0}{r}\right)^{1+\frac{1}{\alpha}}C_1
A\right)  \,.
\end{eqnarray}
For the particular case of $\alpha=1$, the hypergeometric function
takes the form
\begin{eqnarray}
{\rm hypergeom}\left([1/2,-3/2],[-1/2],-{\frac {r_0^2 \,C_1\,A}{r^
2}}\right)=
      \nonumber  \\
=r^{-3} \left (r^2-2\,r_0^2 \,C_1\,A\right )\sqrt {r^2+r_0^2
\,C_1\,A}    \;,
\end{eqnarray}
so that the form function reduces to
\begin{equation}
b(r)=\frac{8\pi\sqrt{A}}{3}\sqrt{r^2+r_0^2 \,
C_1\,A}\;\left(r^2-2r_0^2 \,C_1\,A\right)+C_2 \,.
\end{equation}
Note that the constant $C_1$, for this case, may be obtained from
Eq. (\ref{C1-constant}), and takes the form
\begin{eqnarray}
C_1=\left[\left(8\pi r_0^2 A\right)^{-2}-A^{-1}\right] \,.
\end{eqnarray}

Taking the radial derivative of the form function, or simply using
Eq. (\ref{formderiv}), and inserting the expression deduced for
$C_1$, we have
\begin{equation}
b'(r)=\frac{8\pi \sqrt{A}r^3}{\sqrt{r^2-r_0^2+(8\pi
r_0^2)^{-2}A^{-1}}}  \,,
\end{equation}
which evaluated at the throat, reduces to $b'(r_0)=(8\pi r_0^2)^2
A$. From the condition $b'(r_0)<1$, we verify once again $A<(8\pi
r_0^2)^{-2}$, which is consistent with the generic wormhole
restriction given by Eq. (\ref{ChapWH-restrict}), with $\alpha=1$.
Note that considering $A=\beta(8\pi r_0^2)^{-2}$, with
$0<\beta<1$, the constant $C_1$ takes the form $C_1=(8\pi
r_0^2)^{2}(1-\beta)/\beta^2$. The constant $C_2$, may be evaluated
from the condition $b(r_0)=r_0$, and is given by
$C_2=2r_0/(3\beta)$. The form function is finally given by the
following expression
\begin{eqnarray}
b(r)&=&\frac{r_0}{3\beta}\Bigg\{\sqrt{\beta
\left(\frac{r}{r_0}\right)^2+(1-\beta)} \times
          \nonumber    \\
&&\times\left[\beta
\left(\frac{r}{r_0}\right)^2-2(1-\beta)\right]+2\Bigg\}\,.
\end{eqnarray}

\begin{figure}[h]
\centering
  \includegraphics[width=2.8in]{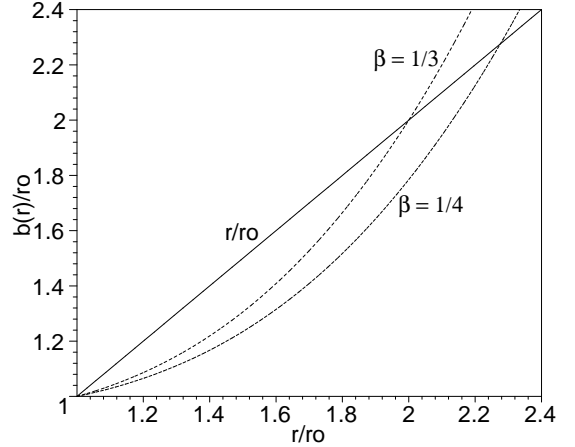}
  \caption{To be a solution of a wormhole, the condition $b(r)<r$ is
  imposed, so that only the interval below the solid line $r/r_0$
  provide wormhole solutions.
  The plot depicts the values of $\beta=1/3$ and $\beta=1/4$, and
  one verifies that the wormhole dimensions increase for decreasing
  values of $\beta$. See the text for details.}
  \label{Fig:form2}
\end{figure}

To be a solution of a wormhole, the condition $b(r)<r$ is imposed,
as emphasized above. This restriction is shown graphically in Fig.
\ref{Fig:form2} for the values of $\beta=1/3$ and $\beta=1/4$. One
verifies that the wormhole dimensions increase for decreasing
values of $\beta$, so that the dimensions of this specific
Chaplygin wormhole cannot be arbitrarily large.

\section{Summary and conclusion}

The generalized Chaplygin gas (GCG) is a possible candidate for
the unification of dark energy, responsible for the present
accelerated cosmic expansion, and of dark matter, inferred, for
instance, from galactic rotation curves. The GCG is parametrized
by an exotic equation of state given by
$p_{ch}=-A/\rho_{ch}^{\alpha}$, where $A$ and $\alpha$ are
positive constants, with $0<\alpha \leq 1$. The GCG models, in a
cosmological context, have at least two significant features that
stand out. Firstly, they describe a smooth transition from a
decelerated expansion of the Universe to a present epoch of a
cosmic acceleration. Secondly, they provide a unified macroscopic
phenomenological description of dark matter and of dark energy.

In this paper, we have studied the possibility that traversable
wormholes may be supported by the GCG. We have found that to be a
generic solution of a wormhole, the GCG equation of state imposes
the following restriction $A<(8\pi r_0^2)^{-(1+\alpha)}$,
consequently violating the NEC condition, which is a necessary
ingredient in wormhole physics. We analyzed the physical
properties and characteristics of these Chaplygin wormholes,
studying in some detail four specific solutions. The first is that
of a constant redshift function, and specific wormhole dimensions,
the traversal velocity and time were deduced from the
traversability conditions for this particular geometry. The second
solution is that of a specific choice for the form function, and
the theoretical construction of this spacetime with infinitesimal
amounts of averaged null energy condition violating Chaplygin gas
was also explored. The third case analyzed is that of a constant
energy density, and finally isotropic pressure Chaplygin
traversable wormhole solutions were also presented. The solutions
found are not asymptotically flat, where the spatial distribution
of the exotic GCG is restricted to the throat vicinity, so that
the dimensions of these Chaplygin wormholes are not arbitrarily
large.

In concluding, it is noteworthy the relative ease with which one
may theoretically construct traversable wormholes with the exotic
fluid equations of state used in cosmology to explain the present
accelerated expansion of the Universe.
As for phantom energy traversable wormholes \cite{Lobo-phantom},
these Chaplygin variations have far-reaching physical and
cosmological implications, namely apart from being used for
interstellar shortcuts, an absurdly advanced civilization may
convert them into time-machines~\cite{mty,Visser,Kluwer}, probably
implying the violation of causality.

\section*{Acknowledgements}

I thank Jos\'{e} Jim\'{e}nez Madrid for pointing out specific
details of the constant $B<0$ and the respective violation of the
dominant energy condition; Neven Bili\'{c} and Xin Zhang for
pointing out several references; and an anonymous referee for the
constructive comments that ameliorated the manuscript.


\end{document}